\documentclass[journal=jctcce,manuscript=article, layout=traditional]{achemso}
\usepackage{helvet}
\usepackage{booktabs}

\usepackage[utf8]{inputenc}
\usepackage[T1]{fontenc}
\usepackage{url}
\usepackage{multirow}
\usepackage{wrapfig}
\usepackage{hyperref}
\usepackage{xcolor, colortbl}
\usepackage{upgreek}
\usepackage{graphicx}

\definecolor{tmavaseda}{gray}{0.2}
\definecolor{svetlaseda}{gray}{0.85}
\hypersetup{
pdfborder={0 0 0 [0 0]},
colorlinks=true,
urlcolor=teal,
linkcolor=tmavaseda,
citecolor=tmavaseda
}

\title{Computer simulations of the bacterial ribosome using a general purpose coarse-grained model MARTINI}

\author{Josef Cikhart}
\affiliation{Department of Physical Chemistry, University of Chemistry and Technology, Technická 5, 16628 Prague, Czech Republic}

\author{Aneta Leskourová}
\affiliation{Department of Physical Chemistry, University of Chemistry and Technology, Technická 5, 16628 Prague, Czech Republic}

\author{Michal H. Kolář}
\affiliation{Department of Physical Chemistry, University of Chemistry and Technology, Technická 5, 16628 Prague, Czech Republic}
\email{michal@mhko.science}

\begin{document}

\setlength{\parindent}{0em}
\setlength{\parskip}{0.7em}

\maketitle

\begin{abstract}
\singlespacing

Ribosomes are critical biomolecular nanomachines responsible for protein synthesis in all known organisms. The function and dynamics of ribosomes can be studied using molecular dynamics computer simulations. Although this task remains challenging at atomic level, several studies have reported all-atom molecular dynamics simulations of the entire ribosome. However, for certain applications, atomistic simulations are impractical due to the limited simulation timescales achievable. In this study, we investigate the applicability of the coarse-grained MARTINI model for simulations of the bacterial ribosome. After testing several simulation setups, we found that the structure of the ribosome and its components are generally well represented compared to the reference experimental structure. Compared with all-atom simulations of the entire ribosome, coarse-grained simulations result in a less flexible and smaller ribosome. We demonstrate how modifications of some parameters of the model can enhance the dynamics of the ribosome to better align with the atomistic model. Our work provides a detailed protocol for coarse-grained simulations of the ribosome and highlights aspects of the model that need improvements.

\end{abstract}

\singlespacing

\section{Introduction}

The ribosome is a fundamental biomolecular machine responsible for the synthesis of all proteins throughout life. Our understanding of ribosome structure and function has improved dramatically since more than half a century ago, when ribosomes were discovered \cite{Palade1955}. A great amount of structural information has been gathered using X-ray crystallography and cryogenic electron microscopy (cryo-EM), as reflected in the Nobel Prizes for Chemistry in 2009 \cite{Ban2000,Schluenzen2000,Wimberly2000} and partially also in 2017~\cite{Frank1995}. However, the dynamic aspects of protein synthesis in ribosomes remain less explored, despite the fact that they play a fundamental role.

Molecular dynamics (MD) simulations are a powerful tool for investigating the dynamics behavior of molecular systems based on classical potential \cite{Hansen2013}. Since the 1970s \cite{McCammon1977}, all-atom MD (aaMD) simulations have become a well-established method to explore the detailed dynamics of various biomolecules \cite{Karplus2002,Lee2009,Hollingsworth2018} and even complex multicomponent systems, such as the ribosome \cite{Bock2018, Bock2023}. However, simulating an entire ribosome remains a formidable challenge for three major reasons. First, the bacterial ribosome in an explicit solvent environment requires a simulation box containing more than two million atoms, with overall dimensions exceeding 30\,nm. The simulation boxes for the eukaryotic ribosomes are even larger. Second, the chemical composition of the ribosome is complex, including canonical and noncannonical nucleotides, folded proteins, and proteins with disordered regions. And third, the timescales relevant for ribosomal function span an extensive range, from picoseconds to seconds. For example, peptidyl transfer from the P-site tRNA to the A-site tRNA occurs with a maximum rate constant of approximately 10--30\,s$^{-1}$, depending on the substrate \cite{Johansson2011}. Capturing such timescales is computationally prohibitive using traditional MD simulations, as the timestep for numerically integrating Newton's equations of motion is typically 1--4 fs.

One strategy for tackling the enormous computational demands of aaMD simulations of the ribosome or other large biomolecular system is to sacrifice atomic resolution and study molecular models with coarser resolution. The thought of coarse-graining in statistical mechanics first came from Gibbs, in his work on the entropy of systems in phase space adhering to the Liouville theorem (conserving phase-space volume) \cite{Robertson2020}.

To study biological systems, three primary approaches to coarse-graining have emerged \cite{Joshi2021}: top-down, bottom-up, and hybrid \cite{Roel-Touris2020}. The top-down approach starts with an experimentally determined biological structure and tunes the force field (FF) parameters to reproduce experimental macroscopic observables, often focusing on thermodynamic metrics such as stability or folding energetics \cite{Cooke2005}. A prominent example of a top-down approach is the G\={o} model \cite{Takada2019}, also known as a structure-based model. This model assumes that the native structure corresponds to a global free-energy minimum, effectively smoothing the free-energy landscape to favor conformations near the experimentally observed structure \cite{Taketomi1975, Hyeon2011}. By biasing the FF parameters towards this free-energy minimum, the G\={o} model has been extensively used to study protein folding dynamics, where the native structure plays a critical role in defining the folding pathway \cite{Hills2009a}.

In contrast, the bottom-up approach, often referred to as the physics-based method, derives FF parameters directly from the thermodynamic and statistical mechanics principles underlying interactions within all-atom systems. In mathematical terms, it can be thought of as a mapping from the all-atom (AA) phase space to the coarse-grain (CG) phase space with a projected Hamiltonian parametrized through an objective function to reproduce the AA system in certain quantities \cite{Noid2013, Jin2022}. Therefore, one should also be able to describe the microscopic structural properties of the fine-resolution system with reasonable accuracy and infer new information about the original system \cite{Noid2013}. For example, Savelyev and Papoian have utilized a bottom-up approach to study the dependence of the persistence length of double-stranded DNA and the ionic strength of the solvent \cite{Savelyev2010}.

Originally developed as a top-down model for coarse-grained lipids and lipid membranes \cite{Marrink2007}, the MARTINI model has evolved into one of the most widely used CG models in biomolecular simulations. The MARTINI force field exemplifies a hybrid approach that integrates top-down and bottom-up methodologies in its design \cite{Marrink2023}. Nonbonded interactions are parameterized on the basis of experimental thermodynamic data, such as partitioning free energies between polar and apolar phases, and are modeled using the computationally efficient Lennard-Jones potential. Effective bonded interactions, on the other hand, are calibrated with respect to their distribution in AA systems, ensuring compatibility with structural features observed at higher resolution \cite{Monticelli2008,Marrink2023}.

Although MARTINI may sacrifice some degree of accuracy compared to purely top-down or species-specific models, it compensates for it with its broad applicability and computational efficiency \cite{Marrink2023}. Its versatility has made it particularly attractive for the study of complex, heterogeneous systems that are common in biophysics, significantly simplifying the setup of such simulations \cite{Souza2021}. MARTINI has been successfully applied to a wide range of biological and biophysical systems, including lipid bilayers, proteins, nucleic acids, and multicomponent systems \cite{Perez-Sanchez2023, Chiariello2024}. Recently, it has even been used to coarse-grain an entire bacterial cell, resulting in a system containing approximately 561 million particles -- CG beads -- in contrast to the more than 6 billion atoms in its corresponding AA representation \cite{Stevens2023}. Such advancements demonstrate the power of the MARTINI in enabling large-scale biomolecular simulations that would otherwise be computationally prohibitive.

A significant challenge in MARTINI coarse-grained biomolecular simulations is the difficulty in preserving the tertiary structure of biomolecules due to alterations in the free-energy landscape. To address this issue, an additional potential called \emph{elastic network} is often employed. This network consists of a system of harmonic oscillators that connect selected CG beads within a specified cutoff radius, forming a stabilizing scaffold that drives the tertiary structure throughout the simulation towards a reference structure. 

MARTINI software also offers a refinement of this approach, specifically designed for proteins, which is called the elastic network dynamic model (ELNEDIN or sometimes also ElNeDyn) \cite{Periole2009}. The key distinction of ELNEDIN from a simple MARTINI elastic network lies in the mapping strategy for the coarse-grained model. Although conventional methods map the backbone bead to the center of mass of the C$_\alpha$, N, O, and carboxyl group C atoms, the ELNEDIN model maps the backbone bead directly to the C$_\alpha$ atom position. This ensures that the global elastic network maintains the tertiary structure of the initial high-resolution model more accurately \cite{Periole2009}.

MARTINI can be used to study nucleic acids such as DNA double-helix and RNA aptamers \cite{Uusitalo2015}. Currently, the CG bead parameters and the routines used to convert from atomistic to coarse-grained models are compatible with version\,2 of the MARTINI framework. Through 100 ns-long cgMD simulations, Uusitalo \emph{et al.} demonstrated the ability of the model to maintain the tertiary structure of the ribosome \cite{Uusitalo2015}. In our work, we expand on their preliminary findings and explore them in more detail. We conducted a thorough investigation of the feasibility and methodology of using MARTINI to simulate the bacterial ribosome. Specifically, we examine the impact of elastic networks on the structure and dynamics of the CG ribosome, using microsecond time-scale aaMD simulations and cryo-EM structure as our reference points.

\section{Methods}

\subsection{Simulated system}

The structure of an \emph{E.\,coli} ribosome was derived from a cryo-EM model with a resolution of 2.9\,\AA (PDB: 5AFI \cite{Fischer2015}). The structure was completed to account for the flexible parts missing in the experimental model as described in our previous work \cite{McGrath2022}. The EF-G/tRNA complex was removed so only the tRNA in the peptide site of the ribosome was considered. Then, we replaced the non-cannonical nucleotides found within the experimental model by their chemically closest canonical counterparts to remain consistent with the available AA-to-CG mapping routines of the MARTINI toolbox \cite{deJong2013}.

The CG ribosome was then placed in a periodic rhombic dodecahedron box with a minimum distance of 1.2\,nm from the box faces. The CG ribosome was energy minimized in a 50-step vacuum simulation. The model was then solvated using MARTINI CG water beads \cite{Marrink2007}. The system was neutralized by replacing the CG water beads with beads representing hydrated sodium and chloride ions. The concentration of these hydrated ion beads was set at 0.01\,$\mathrm{mol\,l^{-1}}$. No structural magnesium cations determined by cryo-EM were used. The final simulation box contained about 170 thousand particles.

The reference AA trajectories were taken from our previous study \cite{McGrath2022}. In summary, the simulations involved the entire \emph{E.\,coli} ribosome in explicit water. Amber family of force fields was used to describe the ribosome together with the SPC/E water model, as detailed in the original study \cite{McGrath2022}. Each of the four independent AA trajectories was 1\,$\mu$s long.

\subsection{Coarse-grained potential}

For ribosomal proteins (r-proteins), MARTINI 2.6 was used with the python program \texttt{marti\-nize.py} \cite{deJong2013}. We used the default bonded, angular, dihedral, and improper dihedral harmonic potential parameters. The non-bonded interactions of the non-polar particles were described by the Lennard-Jones potential with the default parameter $\sigma$: $\sigma_{\mathrm{normal}}=0.47$\,nm, $\sigma_\mathrm{S}=0.43$\,nm and $\sigma_\mathrm{T}=0.32$\,nm, for particles of normal size, small (S) and tiny (T), respectively. Interactions between charged (Q) particles were described by the Coulomb potential. 

Ribosomal RNA (rRNA) was modeled by MARTINI 2.2 \cite{deJong2013}, as described in \cite{Uusitalo2015}. In this model, about 6--7 atoms in the AA model are mapped onto a single bead. Although rRNA is single-stranded in solution, in the ribosome, it forms many loops and helices. Therefore, we used the dsRNA mapping approach \cite{Uusitalo2015}. 

One key aspect of the formation of base pairs is hydrogen bonding between the nucleobases. However, the MARTINI model does not adequately stabilize hydrogen bonds. To preserve the tertiary structure, an elastic network was incorporated into the simulated system, as previously proposed \cite{Uusitalo2015}. This network was applied separately to each ribosomal component, ensuring that no elastic bonds were formed between different r-proteins or between r-proteins and rRNA.

In this study, four different elastic networks were tested, denoted STIFF, SOFT, SOFT2, and LIMP. Each network has three characteristics: i) the force constant $K_{\mathrm{EN}}$ that describes the stiffness/softness of the network, ii) distance cutoff parameter $R_\mathrm{EN}$, which defines the elastic bonds, and iii) the types of beads involved in the network. The characteristics are summarized in Tab.\,\ref{tab:params}. The parameters for r-proteins in the elastic networks STIFF, SOFT, and SOFT2 were obtained from the proposed interval of values for the ELNEDIN model \cite{Periole2009}.

\begin{table}[tb]
    \centering
    \caption{Overview of elastic network settings. We generated and analyzed 4 independent cgMD trajetories per elastic network, 1000 ns-long each.}
    \begin{tabular}{lllll} 
    \toprule
    Network
    & Target
    & $K_\mathrm{EN}\, [\mathrm{kJ\,mol^{-1}\,nm^{-2}}]$
    & $R_\mathrm{EN}\, [\mathrm{nm}]$
    & Beads involved\\
    \midrule
    \multirow{2}{4em}{STIFF}
    & r-protein
    & 500
    & 0.9
    & BB \\ 
    & rRNA
    & 500
    & 1.0
    & BB1--BB3, SC1--SC4 \\
    \midrule
    \multirow{2}{4em}{SOFT}
    & r-protein
    & 500
    & 0.8
    & BB\\
    & rRNA
    & 13
    & 1.2
    & BB1--BB3, SC1\\
    \midrule
    \multirow{2}{4em}{SOFT2}
    & r-protein
    & 500
    & 0.8
    & BB\\
    & rRNA
    & 13
    & 1.0
    & BB1--BB3, SC1--SC4\\
    \midrule
    \multirow{2}{4em}{LIMP}
    & r-protein
    & 500
    & 0.7
    & BB\\
    & rRNA
    & 13
    & 0.5
    & BB1--BB3, SC1--SC4\\
    \bottomrule
    \label{tab:params}
    \end{tabular}
\end{table}

The STIFF elastic network \cite{Uusitalo2015} had a force constant $K_\mathrm{EN}$ of $500\,\mathrm{kJ\,mol^{-1}\,nm^{-2}}$ for both r-proteins and rRNA. The cutoff distance $R_\mathrm{C}$ was 0.9\,nm for r-proteins and 1.0\,nm for rRNA. The STIFF network involved the rRNA backbone beads BB1--BB3, SC1--SC4 sidechain rRNA beads, and BB beads of r-proteins. 

The SOFT \cite{Uusitalo2015} elastic network for r-proteins was the same as the STIFF network, with the exception of the smaller $R_\mathrm{EN}$ of 0.8\,nm. For rRNA, we reduced $K_{EN}$ to 13\,$\mathrm{kJ\,mol^{-1}\,nm^{-2}}$ and increased $R_\mathrm{EN}$ to 1.2\,nm. The SOFT network for rRNA involved BB1--BB3 backbone beads and sidechain beads closest to the SC1 backbone. The SOFT2 elastic network for r-proteins was the same as the SOFT network. $R_\mathrm{EN}$ for rRNA was reduced to 1.0\,nm compared to the SOFT network. The rRNA network involved BB1--BB3 backbone beads and all SC1--SC4 sidechain beads, similar to the STIFF network. Finally, we tested presumably the weakest elastic network, designed in this work and denoted LIMP. $R_\mathrm{EN}$ was 0.7\,nm and 0.5\,nm for r-protein and rRNA, respectively. The force constants and the types of beads involved in the LIMP network were kept the same as for the SOFT and SOFT2. For comparison, we performed cgMD simulations without any elastic network, here denoted NONE. 

\subsection{Coarse-grained MD simulations}

The potential energy of the solvated systems was minimized using the steepest descent algorithm until the maximum force was less than 1\,kJ\,mol$^{-1}$\,nm$^{-1}$ or the computer precision was reached. Then, each system was heated to 300\,K for 10 ns with a time step of 5 fs using the v-rescale thermostat \cite{Bussi2007} with a constant volume of the simulation box. To prevent large conformational changes during heating, the Cartesian coordinates of the ribosome were restrained by a harmonic potential with a force constant of 500\,kJ\,mol$^{-1}$\,nm$^{-2}$. The initial velocities were taken randomly from the Maxwell--Boltzmann distribution at 300\,K. Subsequently, to equilibrate the volume and density of the simulation boxes, the systems were subjected to an unrestrained 50\,ns NPT simulation at 1\,bar and 300\,K using the Berendsen barostat \cite{Berendsen1984} and the v-rescale thermostat, respectively.

The production cgMD simulations were performed using the isotropic Parrinello--Rahman barostat \cite{Parrinello1981} with a time constant $\tau_p$ of 12\,ps and the v-rescale thermostat with a time constant $\tau_t$ of 2\,ps. Production simulations were performed using a time step of 5\,fs for a duration of 1000\,ns without any position restraints. 

Each cgMD simulation was run using the leap-frog integration algorithm. The Verlet scheme on a grid was used as a neighbor search algorithm for short-range interactions~\cite{Verlet1967} with the buffer tolerance set to 0.005\,kJ\,mol$^{-1}$\,ps$^{-1}$ and the neighbor list was updated every 30 steps. Reaction field electrostatics was used for the description of Coulomb forces with the relative dielectric constant set $\epsilon_{r}$ to 15 and to infinity outside the cutoff distance of 1.1\,nm, according to other MARTINI simulations \cite{Marrink2023}. The Lennard-Jones potential was cut off and shifted, so it would reach a value of zero at the cutoff length of 1.1\,nm and would be continuous. The minimum cell size due to bonded interactions during domain decomposition was 1.6\,nm. The P-LINCS algorithm of order 4 was used to constrain the length of the bonds \cite{Hess1997}. The ribosome conformations were stored every 10000 steps, resulting in trajectories of 20000 frames.

All production simulations were performed in GROMACS \cite{Abraham2015} 2021.4.

\subsection{Analyses}

We computed various metrics to analyze both the structural and dynamic aspects of the ribosome and its individual parts. We compared results from the cgMD and aaMD simulations as well as between the cgMD simulations and the reference experimental model (PDB 5AFI \cite{Fischer2015}).

Root-mean-squared deviation RMSD was calculated for each trajectory frame $f$:

\begin{equation}
\mathrm{RMSD}(f) = \sqrt{\frac{1}{N_p} \sum_{i=1}^{N_p} |\mathbf{x}_i (f)-\mathbf{x}_i^\mathrm{ref}|^2},
\end{equation}
where $\mathbf{x}_i$ is the position vector of the particle $i$ in frame $f$, $\mathbf{x}^\mathrm{ref}$ is the position vector of the particle $i$ in a reference structure, and $||$ represents the Euclidean norm. The sum runs through $N_p$ particles of the system. 

Root-mean-square fluctuation RMSF was calculated for each bead $i$:

\begin{equation}
\mathrm{RMSF}(i) = \sqrt{\frac{1}{N_f}\sum_{f=1}^{N_f}|\mathbf{x}_f(i)- \langle \mathbf{x}(i) \rangle|^2},
\end{equation}
where $\mathbf{x}_{i, f}$ is the position vector of the particle $i$ in frame $f$, $\langle \rangle$ represent the mean value over trajectory frames. The sum runs trough $N_f$ trajectory frames. Here, RMSF was calculated after superposition of C$_\alpha$ and P beads with respect to the reference structure. The average position vectors were calculated from the frames in the second half of the production trajectories. The RMSF values were averaged over the four replicas. Furthermore, RMSF was calculated for each chain (r-protein or rRNA) by averaging the RMSFs of BB and BB1 beads that are equivalent to CA and P atoms, respectively.

To characterize the compactness of the ribosome, the radius of gyration $R_\mathrm{gyr}$ was calculated for each trajectory frame $f$:

\begin{equation}
R_\mathrm{gyr}(f) = \sqrt{\frac{\sum_{i=1}^N r^2_i m_i}{\sum_{i=1}^N m_i}},
\end{equation}
where $r_i$ is the Euclidean distance between the particle $i$ and the center of mass of $N$ selected particles of weights $m_i$. Since the coarse-grained beads of CA and P atoms weigh the same, the masses were set to 1 to focus solely on the geometric size.

To characterize the solvation of the ribosome, we calculated the number of water beads in the volume of the ribosome. We considered all waters closer to 0.8\,nm from any ribosome beads. 

\begin{figure}[h]
    \centering
    \includegraphics[scale=0.9]{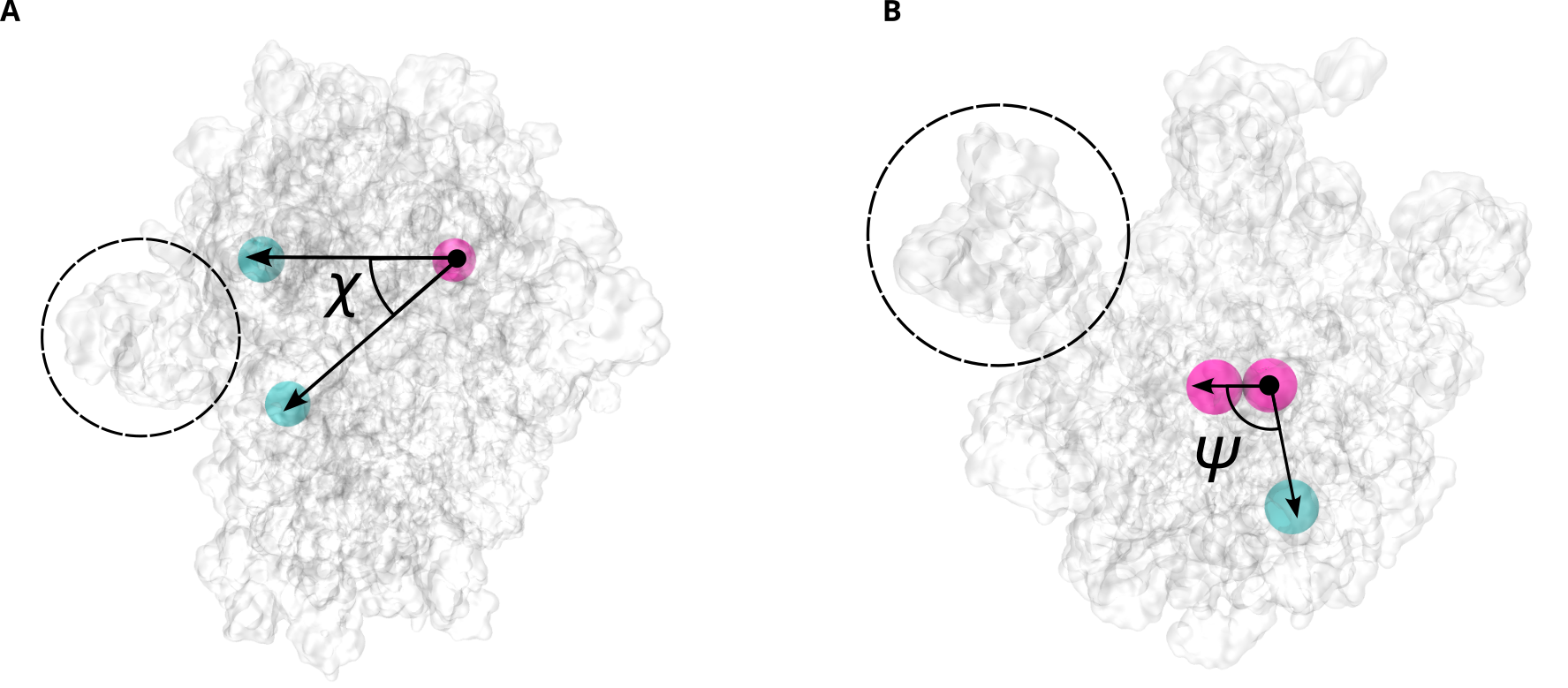}
    \caption{A) Schematic representation of the angle $\chi$. The positions of the points R and T$_1$ in cyan were updated throughout the analysis. The position of the point R in magenta was kept constant from the starting frame of each trajectory. B) Schematic representation of the angle $\psi$ in crown view of the LSU. The position of the point T$_3$ in cyan was updated throughout the analysis. The position of the points $\mathrm{R}_1$ and $\mathrm{R}_2$ in magenta was kept constant from the starting frame of each trajectory. In both A) and B), the L1 stalk is highlighted to better show the position of the points relative to the ribosome.}
    \label{fig:angle_schematic}
\end{figure}

We analyzed large-scale movements of large subunit (LSU) and small subunit (SSU) by computing the intersubunit angles $\chi$ and $\psi$ (Fig.\,\ref{fig:angle_schematic}). The trajectories were aligned with respect to the C$_\alpha$ and P beads of SSU. For $\chi$, three points were defined as the centers of mass of selected residues: A reference fixed point R in LSU 23S rRNA, a point P in the SSU 16S rRNA test point T$_1$ in the LSU by residue 17 of 23S rRNA. Then, the vectors $\overrightarrow{RP}$, $\overrightarrow{RT_1}$ were defined. $\chi$ was defined to capture motion on a different axis. Two fixed reference points $\mathrm{R}_1$ by residue 1940 and $\mathrm{R}_2$ by residue 1903 and a test point T$_3$ by residue 1682 in the LSU, all in 23S rRNA. The angle was then measured between vectors \overrightarrow{R_1T_3} and \overrightarrow{R_1R_2}.

To visualize the distribution of RMSF data, we employed the Empirical Cumulative Distribution Function (ECDF), as implemented in the Seaborn \cite{Waskom2021} statistical data visualization library. The ECDF provides a non-parametric estimate of the cumulative distribution function of a random variable based on the observed data.

Given a sample $\{x_1, x_2, \ldots, x_n\}$ of size $n$, the ECDF $F_n(x)$ is defined as:
\begin{equation}
    F_n(x) = \frac{1}{n} \sum_{i=1}^{n} \mathbf{1}_{\{x_i \leq x\}},
\end{equation}

where $\mathbf{1}_{\{x_i \leq x\}}$ is the indicator function, which equals 1 if $x_i \leq x$ and 0 otherwise. This function increases in steps of $1/n$ at each data point and provides a complete view of the distribution without requiring binning, unlike histograms. In our analysis, ECDFs were plotted using Seaborn’s \texttt{ecdfplot()} function, which sorts the sample data and returns the proportion of observations less than or equal to each unique value.

To measure the occurrence of intersubunit motion, we calculated the change in RMSD depending on least-square-fit alignment. In the rist place, we aligned the CA and P beads of the ribosome and then aligned on the CA and P beads in the SSU only.

For analyses and processing of our production run data, we used in-house Python scripts with the NumPy \cite{Harris2020}, SciPy \cite{Virtanen2020}, Seaborn \cite{Waskom2021} and MDAnalysis \cite{Gowers2016, Michaud-Agrawal2011} libraries and internal GROMACS tools \cite{Abraham2015}. For visualization purposes, the VMD software was used \cite{Humphrey1996}.

\section{Results and Discussion}

\subsection{Differences between the elastic networks}

\begin{figure}[tb]
    \includegraphics[width=\textwidth]{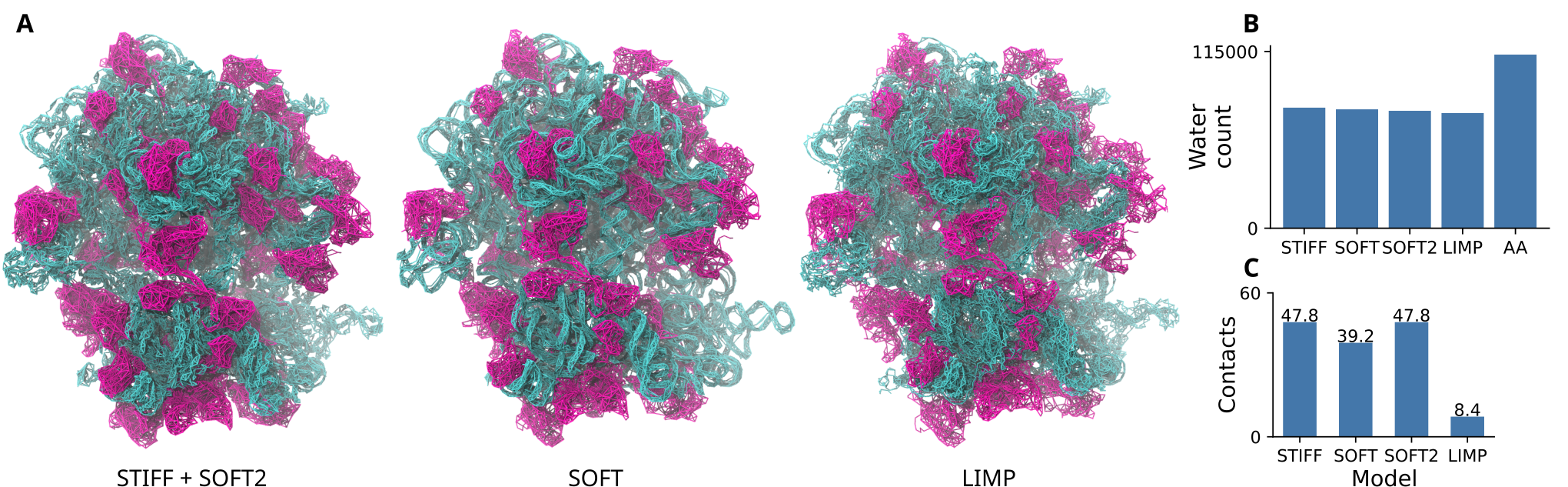}
    \caption{A) Visualizations of the \textit{E.\,coli} ribosome in each of the elastic network models with rRNA in cyan and r-protein in magenta. B) number of water molecules inside the structure in each of the elastic network models. C) Average number of elastic network connections in the rRNA.    }
    \label{fig:structure}
\end{figure}

Fig.\,\ref{fig:structure} summarizes the elastic networks used in this work to support the tertiary structure of the ribosomes. Upon visual inspection, the differences between elastic networks may at first appear small. However, we can see that the number of contacts between beads varies based on the cutoff length and the types of beads involved in the network (Fig.\,\ref{fig:structure}C). The STIFF and SOFT2 networks have the same number of contacts by definition. The LIMP network is different from the other networks because it has only about 8 contacts on average per rRNA bead. The STIFF, SOFT, and SOFT2 networks have between 39 and 48 contacts per bead. 

Water plays an important role in biomolecular dynamics \cite{Laage2017}. In fact, the ribosome is highly solvated. Water and dissolved ions are required to compensate for the negative charges of the phosphate groups of the rRNA, although some negative charges are screened by the basic side chains of r-proteins. An analysis of the atomistic simulation model of the \emph{E.\,coli} ribosome reveals that the ribosome contains about 33 mass\% of water in its volume organized in many small channels, including the main void of the large ribosomal subunit, the exit tunnel \cite{Samatova2024,Kolar2024}. 

A water CG bead in the MARTINI model represents four water molecules. Its Lennard-Jones $\sigma_\mathrm{W}$ parameter is 0.47\,nm \cite{Marrink2007}, while the oxygen parameter of the water in the SPC/E atomistic water model is 0.3166\,nm \cite{Berendsen1987}. Hence, the CG ribosome is less solvated. We counted water CG beads within the ribosome and compared them with the number of water molecules in the atomistic model (Fig.\ref{fig:structure}). The final frames after 1000\,ns cgMD and aaMD simulations were considered. Taking into account the factor of 4 for the MARTINI water, the CG ribosome contained almost 80 thousand water molecules. The AA ribosome contained more than 110 thousand water molecules. We observed small variations in the number of water molecules among the elastic networks. The CG ribosome with the STIFF network contains by about 5\% more water molecules than the LIMP network.

\subsection{CG ribosome is smaller than AA ribosome}

\begin{figure}[tb]
    \centering
    \includegraphics[scale=0.80]{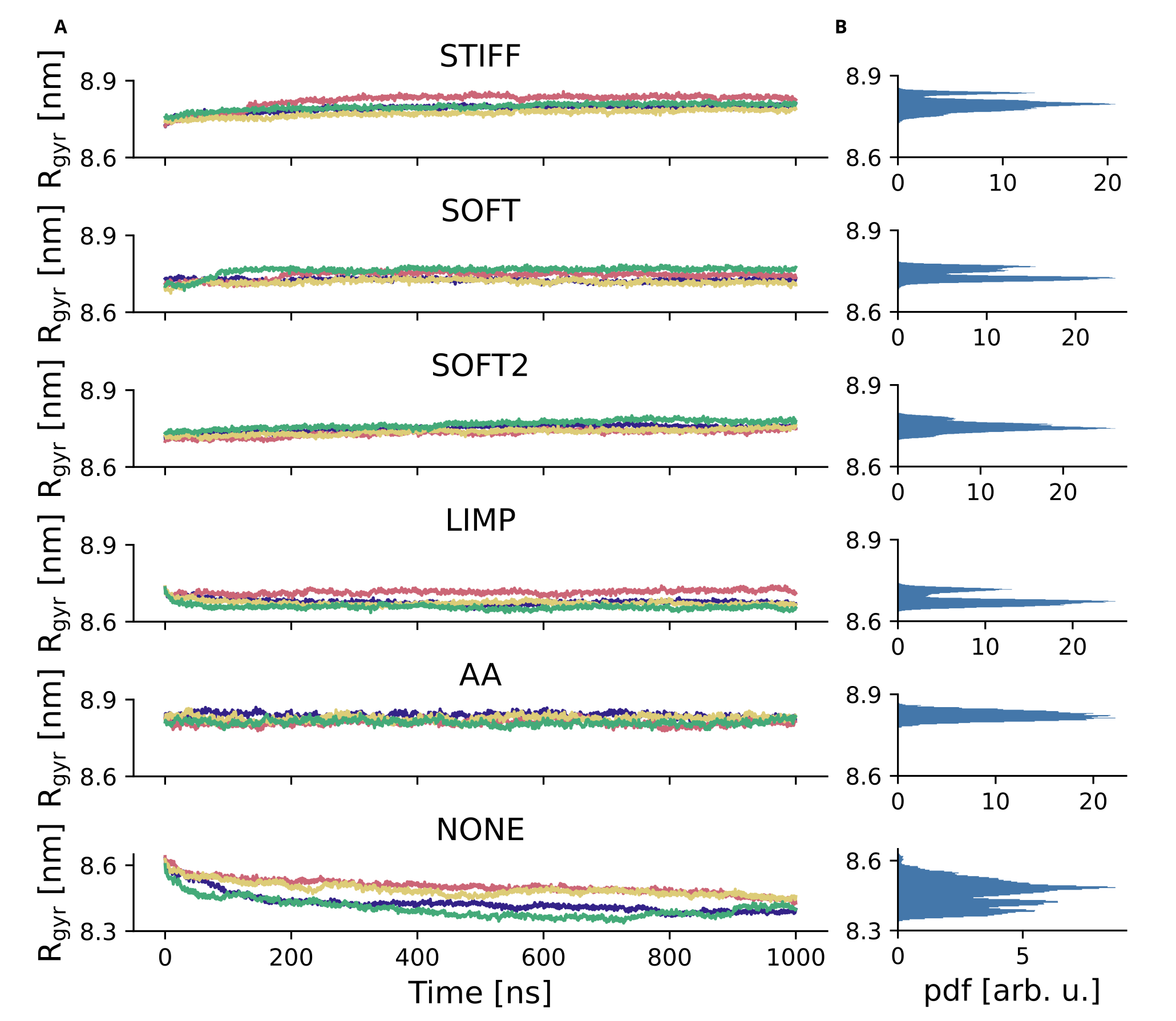}
    \caption{A) R$_\mathrm{gyr}$ as a function of time from each of the trajectories. B) Histograms of R$_\mathrm{gyr}$ values constructed from compound data in A).}
    \label{fig:rgyr}
\end{figure}

For each elastic network, we performed four independent cgMD simulations of 1\,$\mu$s each and compared them to four independent aaMD simulations performed previously in our group for another project \cite{McGrath2022}. The $R_\mathrm{gyr}$ is shown as a function of time in Fig.\,\ref{fig:rgyr} with the respective estimates of the probability density functions (pdf$^{R_\mathrm{gyr}}$). In all cases, the ribosome remained stable and did not unfold. We observed that the size of the ribosome, as characterized by $R_\mathrm{gyr}$, depends on the elastic network. All networks tested produced smaller ribosomes compared to the aaMD model (Fig.\,\ref{fig:rgyr}). The differences were not large, but still significant. The smallest ribosome was obtained by a cgMD simulation with the LIMP network, which was the least rigid model we tested. In fact, we also tested a model without any elastic network (NONE in Fig.\,\ref{fig:rgyr}). The ribosome remained compact and collapsed even more. We conclude that the elastic network in the ribosome is repulsive. It prevents the ribosome from adopting a too compact structure. Similar trends have been described previously by Uusitalo et\,al. \cite{Uusitalo2017} for short nucleic acids. They showed that $R_\mathrm{gyr}$ of MARTINI RNA and DNA is systematically lower than the atomistic counterparts. 

In the coarse-grained simulations of the STIFF, SOFT and LIMP elastic network, a minor, yet clear division of pdfs into two peaks occurred. To understand this bimodality, we first extracted the frames with the top and bottom 5\% $R_\mathrm{gyr}$ values. We then averaged these structures (further referenced as the top and bottom structure) and calculated the RMSD for each chain comprising the ribosome. The three highest RMSD chains are pictured in Fig.\,\ref{fig:motion} A). R-proteins comprising the L1 stalk, the central protuberance (CP), the L7/L12 stalk \cite{Moore2000,Trabuco2010} showed a great amount of movement across all systems in agreement with the aaMD model. Chain 7 of the L1 stalk in particular was the first or second highest RMSD chain between two extracted structures in all simulations. The CG simulations exhibiting a second, higher $R_\mathrm{gyr}$ differed from the aaMD model in this analysis notably in surface r-protein bL9. Nonetheless, apart from the lower RMSD in CG systems, the overall shape of the distribution remained similar to the aaMD simulations, without any

Fig.\,\ref{fig:motion}C shows the complete trajectory of each system's LSU in crown view, with a color gradient that transitions from red (initial positions) through white to blue (final positions). A structural alignment was performed on the C-$\upalpha$ and P atoms (or equivalent beads) of the LSU to emphasize the dynamic motions of key regions. This analysis visually confirmed that all of the CG models studied capture the essential dynamic behavior of the surface ribosomal regions, exhibiting a very good match with the corresponding AA simulations.

\subsection{cgMD ribosome simulations capture intersubunit motion}

\begin{figure}[tb]
    \centering
    \includegraphics[scale=0.90]{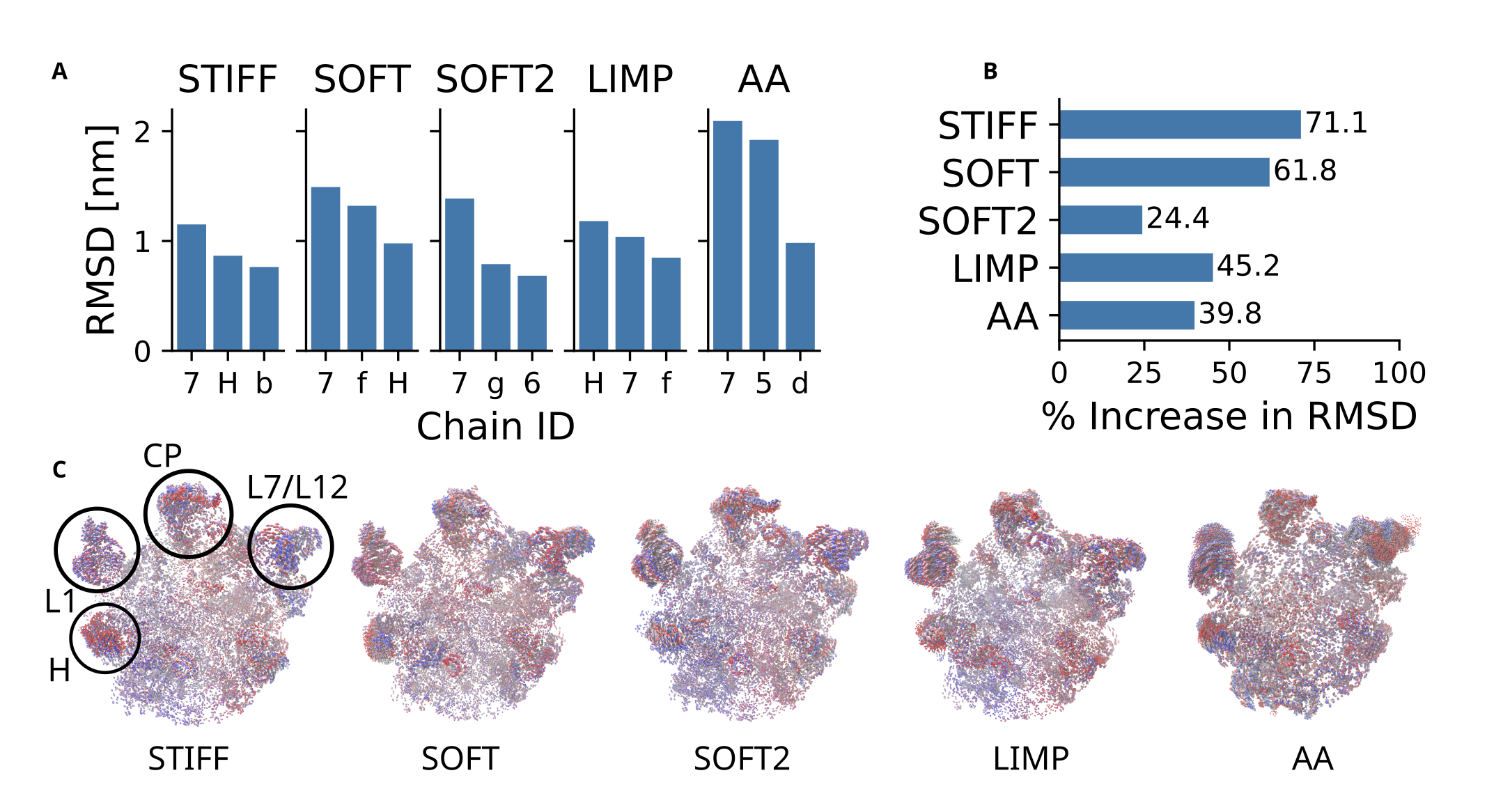}
    \caption{A) The 3 chains with highest RMSD when comparing the top and bottom 5\% $^{R_\mathrm{gyr}}$ structures.  B) C) The crown view of the large subunit with the superposition of the cgMD trajectory frames, evolving in time from red through white to blue. Key regions for the mobility of the ribosome are highlighted.}
    \label{fig:motion}
\end{figure}

To investigate the bimodality of pdf$^{R_\mathrm{gyr}}$ deeper and link them to a possible intersubunit motion, we extracted the frames with the top and bottom 5\% $R_\mathrm{gyr}$ values. We then averaged them and calculated the increase in RMSD between them based on alignment, as shown in Fig.\, \ref{fig:motion}B. We saw the most substantial increase in the STIFF elastic network, followed by the SOFT and LIMP elastic networks. This data indicated that the second peak in pdf$^{R_\mathrm{gyr}}$ could plausibly be caused by intersubunit motion. During visual inspection, we indeed saw that, apart from the movement of surface r-proteins, large-scale motion occurred in the higher $R_\mathrm{gyr}$ trajectories.

\begin{figure}[tb]
    \begin{center}
    \includegraphics[scale=0.79]{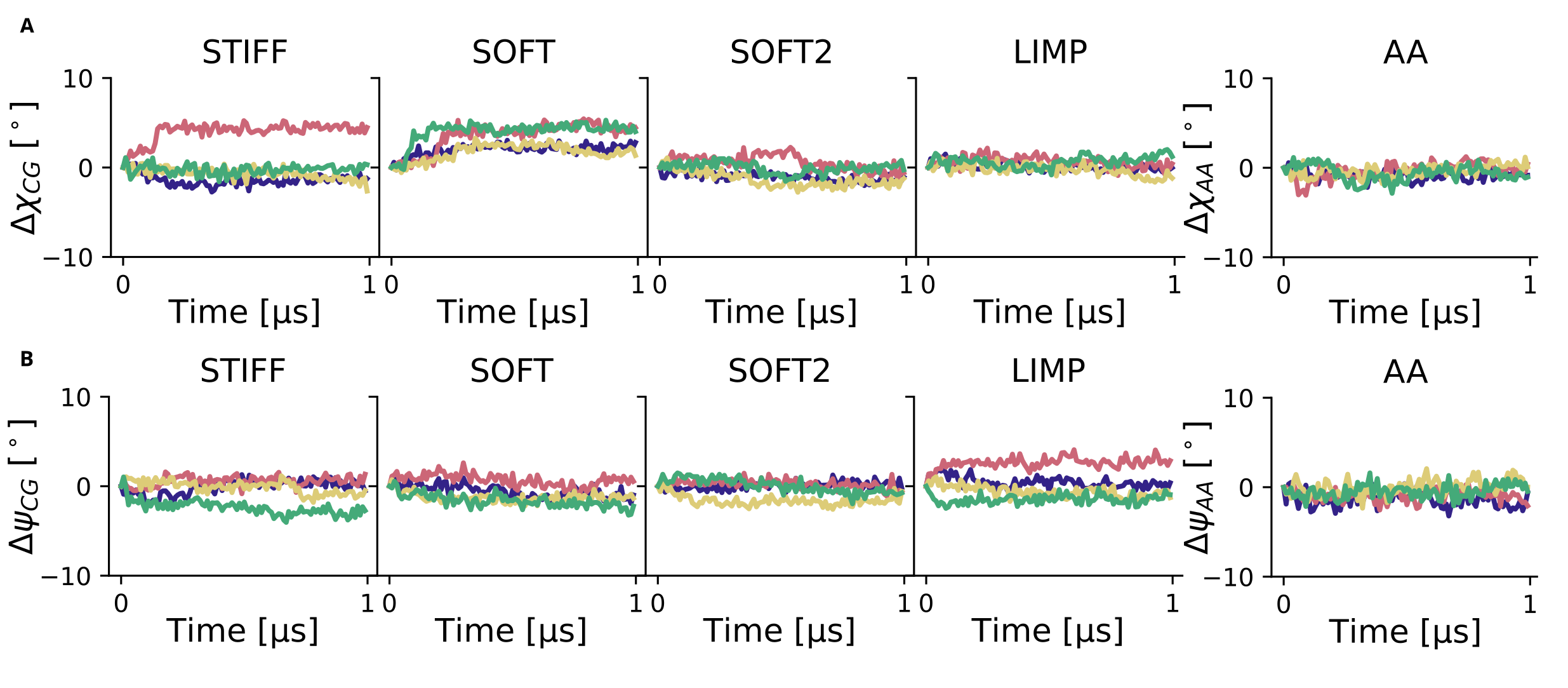}
    \caption{A)Difference between the angle $\chi$ in frame 0 and current frame angle $\chi$, marked as $\mathrm{\Delta}\chi$ as a function of simulation time.}
    \label{fig:rotation}
    \end{center}
\end{figure}

Fig.\,\ref{fig:rotation} shows the difference in angle $\chi$ and $\psi$ in each frame for four replicas of each system. It is apparent that the STIFF elastic network underwent a transition in one of the four trajectories into the higher $\chi$ state and remained in its free energy minimum for the entire trajectory. In the SOFT elastic network setting, the LSU seemingly performed the observed movement in two replicas of the simulation, however, showing more nuance to its transitions rather than a single moved state in one trajectory. Neither of these elastic network settings observed a return to the original state. The SOFT2 elastic network exhibited the least movement among all the systems as defined by angles $\chi$ and $\psi$. The LIMP elastic network showed a clear change in angle $\chi$, unlike the other networks. The difference was most apparent in one replica. The AA model showed no transition between states based on the angles $\chi$ or $\psi$.

\subsection{CG ribosome is less flexible than AA ribosome}

\begin{figure}[ht]
    \begin{center}
    \includegraphics[scale=1.1]{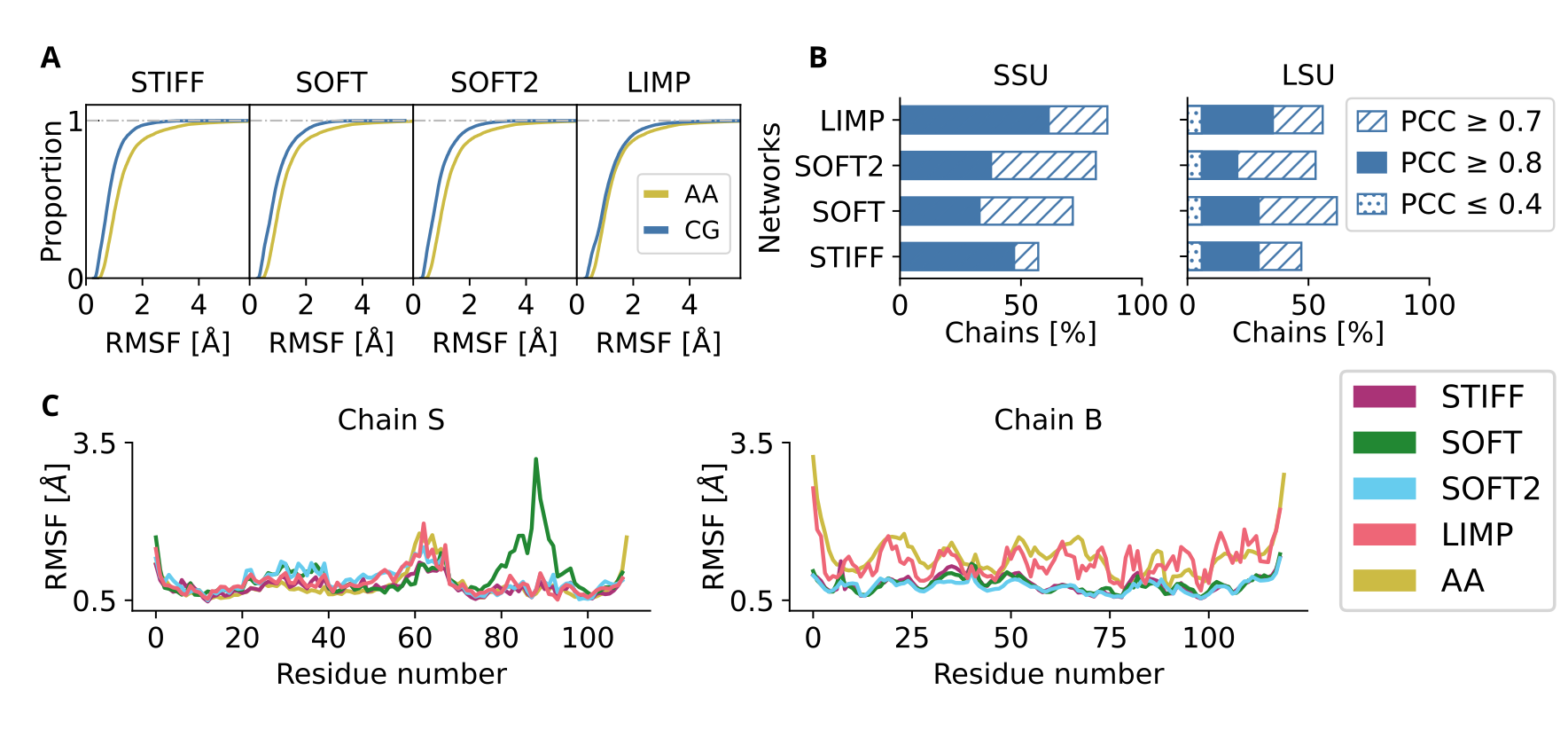}
    \caption{A) Empirical cumulative distribution functions of the RMSF values of each residue in the ribosome. B) Percentage of chains with a Pearson correlation coefficient (PCC) value above 0.7, above 0.8 and below 0.4 in the SSU and in the LSU. C) Per-residue RMSF values of each of the chains comprising the ribosome that scored the lowest PCC values across all of the simulated systems.}
    \label{fig:rmsf}
    \end{center}
\end{figure}

RMSF values were calculated for each residue and averaged to get RMSF of individual chains. Fig.\,\ref{fig:rmsf}A presents the empirical cumulative distribution functions (ECDFs) of the RMSF values for all CG residues in blue, compared to their AA counterparts in yellow. The CG models were less flexible than AA model, regardless of the elastic network, as can be seen from the shape of the ECDF profile. Nevertheless, the LIMP elastic network demonstrated the best agreement with the AA model in the cumulative distribution of absolute RMSF values.

To account for the type and position of the residue, Pearson's correlation coefficients (PCCs) were calculated for each r-protein and rRNA chain within the ribosome. The percentage of chains that achieve high correlations (0.8 $\leq$ PCC $\leq$ 1.0) is shown in Fig.\,\ref{fig:rmsf}B for the SSU and the LSU. In the SSU, the LIMP elastic network exhibited the highest correlation, with approximately 62\% of chains displaying PCC values in this range, significantly exceeding the other models. Similarly, in the LSU, the LIMP elastic network achieved the best correlations, with 36\% of the chains falling within the range of (0.8 $\leq$ PCC $\leq$ 1.0). Despite the overall lower PCC values in the LSU, fewer than 6\% of the chains scored a PCC below 0.4. These findings highlight the ability of the LIMP elastic network to capture residue-level dynamics comparably to aaMD simulations, particularly in regions of high mobility.

Furthermore, more than 50\% of the LSU chains in all systems exhibited PCC values within the range of (0.7 $\leq$ PCC $\leq$ 1.0), suggesting a robust correlation for most chains, with a slight decrease near the 0.8 threshold. This trend may be attributed to the previously mentioned highly mobile regions of the LSU. In particular, none of the SSU chains in the LIMP elastic network scored a PCC below 0.4, and more than 85\% of the SSU chains exhibited PCC values in the range of (0.7 $\leq$ PCC $\leq$ 1.0). 

Throughout our simulations, only five chains exhibited a PCC below 0.4 in at least one elastic network setting, all of which were located within the LSU. In 5S rRNA (marked as chain B in Fig.\,\ref{fig:rmsf}), the LIMP elastic network exhibited excellent agreement with the AA model as can be seen in Fig.\,\ref{fig:rmsf}. This contrasts with other elastic network configurations, which consistently underestimated fluctuations across the chain. This again indicates that a less rigid elastic network might capture the fine dynamics of large rRNA molecules the best.

Protein uL22 provided another compelling case, with the SOFT elastic network producing a pronounced RMSF peak between residues 80 and 100. These residues are in located within an extended loop of the r-protein, which is embedded in 23S rRNA and protrudes into the ribosomal exit tunnel. Since elastic networks for RNA were the only variables in these simulations, it is plausible that this motion arises from specific adjustments in the elastic network of RNA 23S rRNA. Remarkably, this pronounced motion was observed in all four SOFT network replicas, further indicating that the RNA elastic network configuration drives this phenomenon. 

\subsection{Analysis of protein--RNA interactions}

\setlength{\belowcaptionskip}{-10pt}
\begin{figure}[ht]
    \begin{center}
    \includegraphics[scale=1.05]{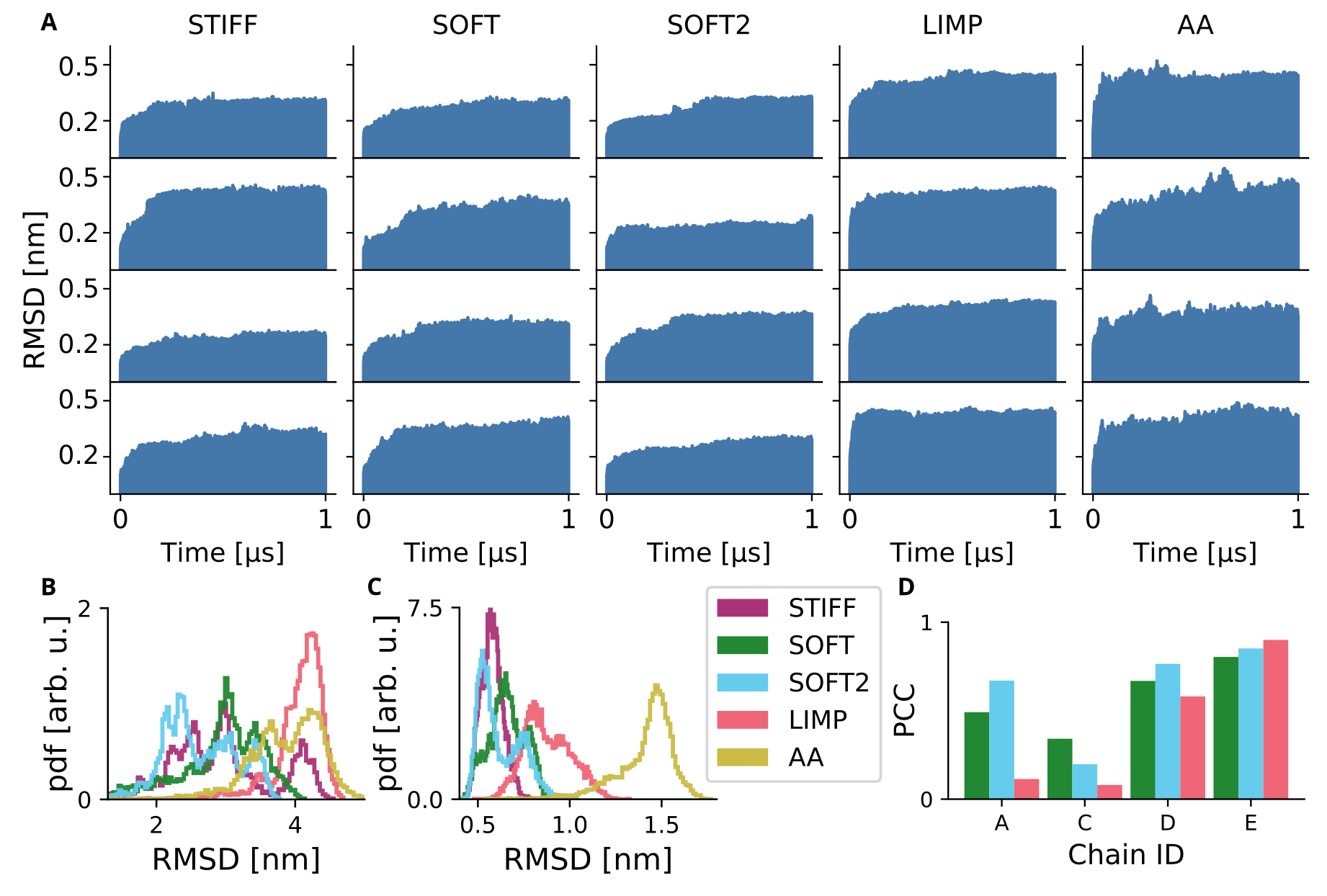}
    \caption{A): Time dependence of the RMSD of the cut-out of the PTC region in all trajectories and systems with an alignment on the P atoms or equivalent CG beads. B) Histograms of RMSD values of the entire ribosome in each of the elastic networks with an alignment done on P atoms or equivalent CG beads. C) Histograms of RMSD values of cut-out around the PTC region of the ribosome in each of the elastic networks with an alignment done on P atoms or equivalent CG beads. D) Pearson correlation coefficients (PCC) of each chain comprising the PTC with the entire PTC RMSD in SOFT, STIFF and LIMP models.}
    \label{fig:rmsd}
    \end{center}
\end{figure}

To examine the ribosome dynamics, the values of the root mean square deviation (RMSD) values were calculated for the C-$\upalpha$ and P atoms in the AA model, or their equivalent beads in the CG model. Structural alignment was performed exclusively on the P atoms or equivalent beads, with a focus on RNA to isolate and better characterize the dynamic interplay between RNA and protein components. This approach reflects the fact that the RNA and protein chains in the CG model are not connected by native bonds nor elastic network bonds. From Fig.\,\ref{fig:rmsd}A we can see that all CG models were stable on the simulation timescale, as also shown in Fig.\,\ref{fig:rgyr} for $R_\mathrm{gyr}$.

Peptidyl transferase center (PTC) is a critical region of the large subunit, where the peptide bond formation occurs. To understand, how the simulation model describes the structure and dynamics of PTC, we selected a sphere with a 2.4\,nm diameter around the center of geometry of residues U2506 and U2586. This region included approximately 100 residues in total. 

Fig.\,\ref{fig:rmsd} shows the histograms of the RMSD values calculated for the PTC. All elastic networks show a maximum in the pdf$^{\mathrm{RMSD}}$ at lower values than for the AA model. This indicates that the CG models were less mobile. We can also see that the LIMP elastic network produced a wider histogram compared to the very narrow peaks of the other CG systems. This is more akin to the AA model and demonstrates that to reproduce the flexibility of the atomistic ribosome, a weaker elastic network setting than STIFF and SOFT is necessary.

Another feature of the pdf$^{\mathrm{RMSD}}$ of the SOFT2, SOFT and LIMP CG models was the bimodal nature of them, indicating transitions between two states. In fact, when we compared the RMSD of the PTC region with the RMSD of each r-protein and 23S rRNA chain (marked chain A in Fig,\ref{fig:rmsd}) whose section overlaps with the PTC region, we found that the PCC between the extended loops of the r-proteins uL3 and uL4 displayed a high correlation. It means that these chains, which also line the ribosomal exit tunnel, are mostly responsible for the overall RMSD of the PTC. Especially in the LIMP elastic network, where the PCC exceeded 0.90 for the interior parts of uL4. However, the SOFT and SOFT2 model showed higher correlations in 23S rRNA as well.

\section{Conclusions}

We have conducted a series of 1,$\upmu$s long simulations of a coarse-gained bacterial ribosome using the MARTINI model for proteins and nucleic acids. The stable timestep in each of the simulations was 5 fs (our test with higher time steps resulted in simulation crashes). Using a single node equipped with two 64-core AMD EPYC 7H12 processors, we achieved a simulation performance of 600--650 ns/day with our CG ribosome consisting of about 171,000 particles including solvent and ion beads. This represents roughly 100-fold speedup compared to the AA simulations. The reduction of the number of particles is the dominant factor.

Our analysis demonstrates that all CG models produced a reasonable ribosome structure, effectively capturing key features of ribosomal dynamics. Among the models, the LIMP elastic network setting exhibited the highest correlation with the AA model in terms of fluctuations and the RMSD of the PTC in the large subunit, and the ribosome as a whole. These results suggest that the LIMP elastic network provides a robust framework for studying the dynamics and movements of both r-protein and rRNA chains over large timescales. On the other hand, it has limitations in regards to the size of the ribosome itself. The choice of elastic network parameters significantly influences the outcome, emphasizing the importance of careful optimization to balance structural stability and physical accuracy.

It is also crucial to consider the impact of coarse graining on the size and solvation of the ribosome. Due to the inherent limitations of CG models, none of our systems achieved the solvation levels observed in the AA simulations. This discrepancy highlights a potential limitation of CG models in the accurate study of solvation effects. However, with properly tuned elastic networks, CG models remain valuable tools for exploring the large-scale dynamics of complex biomolecular assemblies without compromising structural integrity.

\section*{Acknowledgment}

The input simulation data, output data and the analysis scripts used to generate the figures are available online at https://github.org.

\section*{Funding Information}

This work was supported by the Czech Science Foundation (project no. \emph{23-05764S}), the Ministry of Education, Youth and Sports of the Czech Republic through the e-INFRA CZ (ID:90254), and from the grant of Specific university research (A1\_FCHI\_2024\_001).  

\section*{Authors Contribution}

MHK designed and supervised the research, and acquired funding; AL prepared the initial simulation systems and performed the simulations; JC updated the simulated systems, performed simulations, and analyzed them; JC and MHK interpreted the results; JC wrote the initial version of the manuscript; all authors finalized the manuscript.

\section*{Conflict of interest}

The authors declare no conflict of interest.

\providecommand{\latin}[1]{#1}
\makeatletter
\providecommand{\doi}
  {\begingroup\let\do\@makeother\dospecials
  \catcode`\{=1 \catcode`\}=2 \doi@aux}
\providecommand{\doi@aux}[1]{\endgroup\texttt{#1}}
\makeatother
\providecommand*\mcitethebibliography{\thebibliography}
\csname @ifundefined\endcsname{endmcitethebibliography}
  {\let\endmcitethebibliography\endthebibliography}{}

\end{document}